\documentclass[sigconf,authorversion,nonacm]{acmart}

\AtBeginDocument{%
  \providecommand\BibTeX{{%
    \normalfont B\kern-0.5em{\scshape i\kern-0.25em b}\kern-0.8em\TeX}}}

\copyrightyear{2023} 
\acmYear{2023} 
\setcopyright{rightsretained} 
\acmConference[CHI EA '23]{Extended Abstracts of the 2023 CHI Conference on Human Factors in Computing Systems}{April 23--28, 2023}{Hamburg, Germany}
\acmBooktitle{Extended Abstracts of the 2023 CHI Conference on Human Factors in Computing Systems (CHI EA '23), April 23--28, 2023, Hamburg, Germany}\acmDOI{10.1145/3544549.3582733}
\acmISBN{978-1-4503-9422-2/23/04}

\acmConference[alt.chi]{CHI ’22 Extended Abstracts,}{April 23--28, 2023}{Hamburg, Germany}
\acmBooktitle{CHI EA '23: Extended Abstracts of the 2023 CHI Conference on Human Factors in Computing Systems, April 23--28, 2023, Hamburg, Germany} 
\acmPrice{15.00}
\acmISBN{978-1-4503-XXXX-X/18/06}

\usepackage{tipa}
\usepackage{epigraph}
\usepackage{Carrickc,lettrine}

\usepackage{pgfornament}
\newcommand{\fancyOrnament}{\begin{center}\pgfornament[width=.6cm]{3}\end{center}}
\usepackage{wrapfig}

\newcommand{\quoting}[1]{``\emph{#1}''}

\usepackage{tikz}
\usetikzlibrary{shapes,arrows}
\usetikzlibrary{positioning,trees,shapes.geometric, arrows, calc, fit} %shadows

\begin{document}

% ks: Love the first bit but how about for the second bit ...
\title[The Systematic Review-lution]{The Systematic Review-lution: A Manifesto to Promote Rigour and Inclusivity in Research Synthesis}
%for The Place at the Table of Secondary Research and Research Synthesis in HCI and Greater Rigour in its Conduct}

\author{Katja Rogers}
\email{k.s.rogers@uva.nl}
\orcid{0000-0002-5958-3576}
\affiliation{%
  \institution{University of Amsterdam}
  \city{Amsterdam}
  \country{Netherlands}
}

\author{Katie Seaborn}
\email{seaborn.k.aa@m.titech.ac.jp}
\orcid{0000-0002-7812-9096}
\affiliation{%
  \institution{Tokyo Institute of Technology}
  \city{Tokyo}
  \country{Japan}
}

\renewcommand{\shortauthors}{Rogers and Seaborn}


% kr focus: starting a discussion on what rigorous sys reviews should look like in HCI (given the many types of knowledge produced in the field).
% ks focus: primary research, making sure that it is rigorous and suitable for sys reviews

\begin{abstract}
The field of human-computer interaction (HCI) is maturing. Systematic reviews, a staple of many disciplines, play an important and often essential role in how each field contributes to human knowledge. On this prospect, we argue that our meta-level approach to research within HCI needs a revolution. First, we echo previous calls for greater rigour in primary research reporting with a view towards supporting knowledge synthesis in secondary research. Second, we must decide as a community how to carry out systematic review work in light of the many ways that knowledge is produced within HCI (rigour in secondary research methods and epistemological inclusivity). In short, our manifesto is this: we need to develop and make space for an inclusive but rigorous set of standards that supports systematic review work in HCI, through careful consideration of both primary and secondary research methods, expectations, and infrastructure. We call for any and all fellow systematic review-lutionaries to join us.
%A manifesto for the importance of and greater rigour in research synthesis in human-computer interaction. 140 words max.
\end{abstract}



\begin{teaserfigure}
  \includegraphics[ width=\textwidth,trim={0 3cm 0 0}, clip]{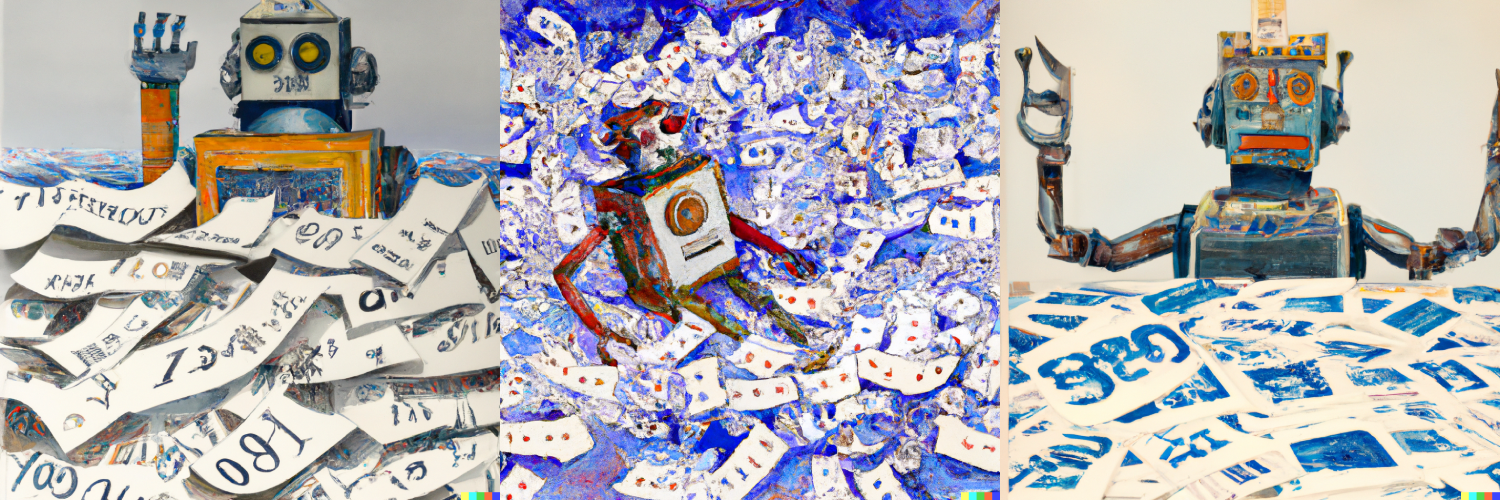} %trim={1cm 10cm 1cm 12cm}, clip,
  \caption{Generated images by Dall-E (\url{https://openai.com/dall-e-2/}) based on the prompt ``a robot drowning among waves of papers, number, and letters, drawn by Monet,'' visualizing the way it sometimes feels to be a researcher in 2022.}
  \Description{Computer-generated expressions of robots being inundated with papers: Three robots absolutely deluged with papers.}
  \label{fig:teaser} 
\end{teaserfigure}


\maketitle

\section*{Publication}

\noindent
The final publication is available via ACM at \url{https://dl.acm.org/doi/10.1145/3544549.3582733}.

\newpage

\section*{Citation}

\noindent
Katja Rogers and Katie Seaborn. 2023. The Systematic Review-lution: A Manifesto to Promote Rigour and Inclusivity in Research Synthesis. In \textit{Extended Abstracts of the 2023 CHI Conference on Human Factors in Computing Systems (CHI EA '23)}. Association for Computing Machinery, New York, NY, USA, Article 424, 1–11. \url{https://doi.org/10.1145/3544549.3582733} \\ \\ \\

%%%%%%%%%%%%%%%%%%%%%%%%%%%%%%%%%%%%%%%%%%%%%%%%%%%%%%%%%%%%%%%%%%
% Instructions: https://chi2023.acm.org/for-authors/alt-chi/

% How well the work embodies alt.chi’s values of creativity, criticality, reflection, and originality.
% How coherent the work is conceptually, philosophically, and methodologically.
% How well the proposed presentation format suits the content of the submission and how likely it is to engage people at the conference.
% The extent to which the work expresses a voice and argument unlikely to be heard in the Papers track.
% The nature of the discussions that the submission provokes among reviewers.

\epigraph{``\emph{We have reason to fear that the multitude of books which grows every day in a prodigious fashion will make the following centuries fall into a state as barbarous as that of the centuries that followed the fall of the Roman Empire.}''}{Adrien Baillet, 1685 (as quoted by Ann Blair) \cite{blair2003reading}}

\section*{Why Research Synthesis Matters}

\begin{figure}%{l}{\columnwidth} %% Answer: [trim = {left, bottom, right, top}, clip]
    \centering
    \includegraphics[trim={0 0cm 1.3cm 0},clip,width=.9\linewidth]{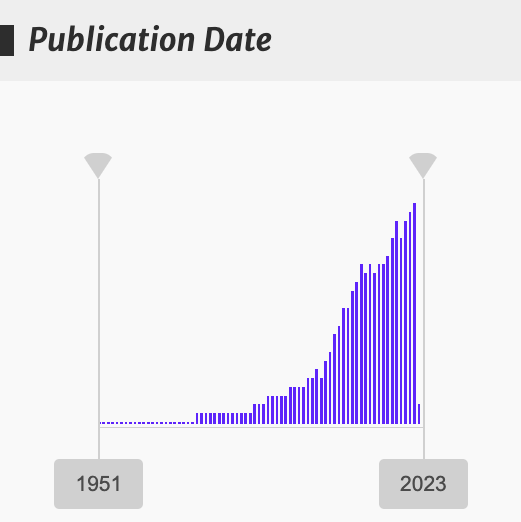}
    \caption{Results per year for the keyword "human-computer interaction" in the ACM Digital Library. A total of 658,883 results were found as of 14:24 on February 7\textsuperscript{th}, 2023. The year 2000 alone featured 8,519 results, with 27,009 for 2010; 35,183 for 2020; and 37,215 results for 2022.} %total results since the year 1951
    %AT SUBMISSION:\caption{Results per year for the keyword "human-computer interaction" in the ACM Digital Library. A total of 642,888 results were found as of 17:00 on October 24\textsuperscript{th}, 2022. The year 2000 alone featured 8,479 results, with 26,985 for 2010, 34,714 for 2020, and as of October, 30,135 results for 2022.} %total results since the year 1951
    \label{fig:hci-growth}
    \Description{A screenshot of the ACM digital library's overview visualisation of search results, ranging from 1951 to 2022, showing a steep increase in publications per year.}
    \vspace{-.3cm}
\end{figure}

\lettrine{T}{h}ose of us in human-computer interaction (HCI) are publishing a lot---dare we say \emph{too much}? This is exemplified by the number of papers in the \href{https://dl.acm.org/action/doSearch?AllField=human-computer+interaction}{ACM Digital Library (DL)} (\autoref{fig:hci-growth}). One person or even a team cannot keep up with the sheer amount of research that we produce. Yet we also need to synthesize this literature.
%, which we often do in the form of systematic reviews. 
In some fields, the enthusiastic pace of research output has led some researchers to explore automation. Modern computing technologies like machine learning  \cite{beller2018making,marshall2019toward,vandeschoot2021open} might be able to aid us in the task of keeping up with and synthesizing publications. But is this necessary or desired in HCI, or do we need to rethink the knowledge production and reporting process?

In other large(r) fields, \citet{chu2021slowed} warn that increased publication output can lead to \quoting{ossification} because novel ideas cannot gain traction against the entrenched canon of the papers most often cited. This can have severe consequences for the field as a whole:  \quoting{too many papers published each year can lead to stagnation rather than advance [knowledge creation]} \cite{chu2021slowed}. There are already hints of this trend in HCI: the papers most cited are cited quite a bit more often than the average paper, while the number of citations papers receive per year is declining overall \cite{pohl2019how}. Based on other fields, this suggests that it is getting more difficult for new ideas to break through and shake up established ones in HCI. 
However, in HCI, we might actually have the opposite problem (too) because of the \emph{kinds} of papers we publish at this rapid pace. In recent years, some researchers have become concerned that we are focusing too much on novelty \cite{balestrini2015civically,oulasvirta2016hci}. Consider the most recent (2022) proceedings of the Conference on Human Factors in Computing Systems (CHI), where searching for "novel" yields 602 results---based on 697 papers. When 86\% of papers are characterized as "novel," we need to ask what this means for knowledge gains and consensus-building. 
Playing devil's advocate, we might say that our field incentivizes, if not requires, the publication of a never-ending stream of flashy one-offs. Instead of putting effort into rigorous incremental research to confirm evidence across multiple studies, we find ourselves dancing after the novelty carrot\footnote{Keeping in mind that our field does not necessarily agree on one meaning of ``novelty.''}.

%Yet, irrespective of the quality of the work produced, only some of us get to take a bite.

The repercussions of this state of affairs deserve careful and critical attention. Focusing so strongly on novelty may be part of what makes it difficult to provide definitive answers about what we actually know so far in HCI \cite{hornbaek2015we}. This also makes it difficult for HCI to situate itself within and participate alongside other fields of study, and limits the kind of research that we do.
We echo \citet{disalvo2010mapping}'s statement on sustainable HCI as relevant for HCI as a whole: \quoting{[t]o avoid reinventing the wheel, there is a need for the field to take stock of what is known and to identify major unknown questions or issues, which arise from what has been established, as a basis for future work.} \citet{whittaker2000lets} raised similar criticisms: that the HCI community \quoting{[overemphasizes] ``radical invention'' at the price of achieving a common research focus.}  They go on to point out that in the absence of \quoting{such a focus, it is difficult to build on previous work, to compare different interaction techniques objectively, and to make progress in developing theory.} This is not merely a problem of research praxis; it also has practical ramifications when we, as a field of study, cannot provide clear guidelines or implications. We find ourselves in a liminal space, where we all are carrying out research and producing a variety of outcomes, but an outsider looking in may find the overall picture difficult to make out. Such an outsider may then move on to a more clearly defined space, leaving our work unacknowledged and overlooked. From within, we may not be able to see the forest for the trees, leaving no clear path forward. 
Research synthesis can clarify the work conducted in a field of study, not only for others but ourselves, as well.

Yet replication studies, follow-up work, corrections and expansions, and other forms of explicitly \emph{not novel} forms of inquiry remain sidelined, despite calls to action that go back more than a decade~\cite{wilson2012replichi,echtler2018open,wilson2011replichi,hornbaek2014is, marshall2017alittle}.  %\cite{disalvo2010mapping}.
This has grave implications in light of larger patterns and hiccups in research practice, including p-hacking \cite{head_extent_2015}, the replication \cite{shrout_psychology_2018, natcomm2022replication} and publication bias \cite{nature_human_behaviour_importance_2019} crises, and adverse effects resulting from the preprint server explosion \cite{noauthor_rise_2020}. 
%Similar concerns have also been echoed in related fields like visualization, where researchers like \citeauthor{correll2022position} are aiming to \quoting{identify [..] struggles
%and opportunities for epistemic reform or growth as a field}~\cite{correll2022position}.
This is not only a problem for experimental or quantitative work typically housed within positivist frameworks.
We recognize that not all research projects within HCI aim for generalization or consensus. Still, many if not all of them hold valuable insights on their own that could be productively synthesized. All epistemological and methodological lenses should be embraced in knowledge synthesis work if we wish to provide a full picture of HCI research. % as a field of study and practice. 
%HRI added replications studies track! also dataset as signposted actual track/submission

Globalized computer-based information technology---the very heart of our discipline---has created new drivers and tensions for scholarship. Solutions to this phenomenon might be found by embracing \emph{slow science} \cite{stengers_another_2018}, to some extent. Yet even if we were to stop publishing altogether tomorrow---and pull the plug on the Internet---we would still need to sift through and synthesize the existing work published so far. Many of us in HCI are taking up this task, but have little guidance or standardization. This is not because guidelines or standards do not exist---they do~\cite{page2021prisma, tong2012enhancing, pullin2006guidelines, snyder2019literature,siddaway2019systematic,haddaway2018roses,kitchenham2007guidelines,topor2020integrative,campbell2020swim}. However, these are premised around the work developed (and valued) in other fields, e.g., randomized-controlled trials in the medical field~\cite{starr2009origins}. % or environmental science~\cite{cee2013guidelines}. 
Part of the challenge inherent to HCI is the \emph{sheer variety} of work available (perhaps even leading to what \citet{reeves2015human} and \citet{fallman2010establishing} refer to as \quoting{disciplinary anxiety}). Closely related to this, another key challenge is the lack of consensus on \emph{how} to carry out research synthesis in general, and systematic reviews more specifically: our field has not yet embarked on an explicit conversation about what we expect from systematic reviews, nor how to handle the different kinds of knowledge our field produces.

This is our manifesto. We propose to begin a community-driven conversation to determine how to depart 
%We propose a solution: a ``novel`` frame to replace the drive for novel research. This will mean, for many, a radical departure 
from our typical research praxis to support research synthesis at the meta level. We argue that a rigorous and inclusive systematic review approach to research synthesis in HCI is the way forward.
We pose two critical questions at this juncture: \emph{1) How can we package our work in such a way that meaningful research synthesis can be practiced based on the wonderful diversity of work that we produce in CHI and adjacent spaces?} 
Secondly, in light of the many forms of knowledge produced in our field, we believe and hope to convince the reader that we must all come together as a community to develop a shared set of research practices for planning, conducting, and reporting research synthesis within HCI: \emph{2) What should research synthesis look like when it is grounded in plurality: quantitative studies, qualitative studies, design research, ethnography, and the development of interactive artifacts and systems?}
%and decide (i) how we can build consensus methodologically within and among the fruits of the CHI cornucopia and (ii) how each of us can prepare our work to meet this end. This is our manifesto. We make one grand assumption: that the greater portion of the CHI community is on board with the disciplinary goal of promoting excellence in research synthesis.
We raise these concerns, challenges, and desires for a different way in the scholarly tradition of denouncing the \quoting{confusing and harmful abundance} of literature, a form of self-reflective discourse that dates back several centuries \cite{blair2003reading}. 
Our goal is to remind ourselves about the bigger picture and re-orient each other as members of a community of practice. %We may need a different approach than previous generations.

With the above, we make the case for research synthesis, why it matters for the field of HCI, and why our answers to this issue may differ from other fields. 
We next 
%In this paper, we lay out a path to achieve the aims of our two-part manifesto. 
%We begin by making the case for research synthesis and why it matters for the field of HCI, and perhaps CHI in particular. 
%We then 
present established methodologies for research synthesis, focusing on the current global standard across most fields of study: the systematic review. We then raise critical issues about relying on a systematic review approach for HCI research, and provocations that anchor to its abundance in topics, methods, and epistemological points of view. We end with a call for action on a novel framing of research synthesis: inviting you, dear reader, and everyone participating in HCI research. We aim to start a conversation at this year's conference that we expect will lead into the development of a future workshop, special interest group, committee and/or collaborations with the goal of establishing a community of practice invested in HCI research synthesis. By gathering the multitude of disciplinary voices and epistemological perspectives in our community, we hope to make a disciplinary impact in terms of knowledge creation and methodology that reverberates back to the larger research community.

%This is where systematic review methods may be able to come to our rescue ... if we make ready for them.

%\katja{need to streamline the intro and make it clearer what already exists / where we are just echoing and boosting criticisms already raised, and where the new arguments begin}

\fancyOrnament

\section*{Launching Pad: What is a Systematic Review? Or: ``But I Have A PRISMA Figure, Surely That Makes It Systematic''}
%\subsubsection*{Or: The Difference Between Literature Review in a Paper or Thesis, and a Systematic Review.}

%\lettrine{O}{r:} \emph{But I Have A PRISMA Figure, Surely That Makes It Systematic.} \\
%\lettrine{O}{r:} \emph{The Difference Between Literature Review in a Paper or Thesis, and a Systematic Review} \\

\lettrine{S}{ystematic} reviews are a staple of research synthesis in many fields of study. In medicine, they are considered \quoting{indispensable} \cite{ioannidis_mass_2016} as the \quoting{gold standard} \cite{moore_systematic_2022} means of arriving at consensus across individual studies on a specific topic, typically an intervention of some kind, so as to enable decision-making grounded in evidence-based work \cite{ioannidis_mass_2016}. Yet there is no clear agreed-on definition for what a systematic review is as an outcome and what it should entail as a process \cite{krnicmartinic2019definition}. As \citet{krnicmartinic2019definition} explain, \quoting{definitions of [systematic reviews] are vague and ambiguous, often using terms such as \emph{clear, explicit} and \emph{systematic}, without further elaboration.} This presents those of us in HCI with a challenge and an opportunity: we may struggle to understand and apply systematic review methodologies to our work, even though we have much to learn from other disciplines in the history and practice of the systematic review. %for our own research synthesis work. 
Yet perhaps we do not need to adopt all of these methods as they are, and in some cases perhaps it would even be inappropriate to try; we may instead chart a new path forward.
%but we may not need to---we may instead chart a new path forward. Nevertheless, we in HCI are standing on the shoulders of giants---or perhaps, with our different lineages, we should say \emph{titans}---and we have much to learn from in the history and practice of the systematic review for our own research synthesis work.

The first step towards understanding and productive deviation is a definition. Let us begin with what a systematic review is not: It is not merely a description of previous work on a certain topic, within a field of study, or around a certain research question or hypothesis. It is not an annotated bibliography in which we comment on papers that we have read. It is also not conducted ad hoc without an \emph{a priori} plan, especially not if the procedure was changed mid-process so that, in the end, the research question had to be adjusted to fit the method. It is not the summary alone; it cannot be without a description of how the results were constructed, magicking its way from search process to outcomes with no in-between. It is not a narrative of a curated selection of works.

Then, what might a systematic review \emph{be}? In other words, what is its nature as an outcome and method of scholarship? We start with a pair of concepts: primary and secondary research---their relationship, and a basic distinction between the two. Primary research is the \emph{material} of a systematic review.
%, and perhaps most reviews. 
We define \emph{primary research} as any paper\footnote{While we use "paper" here, we do not mean that the paper itself is "the research." We use "paper" for simplicity in writing and in recognition that most output of scholarship is packaged in paper form, particularly in the context of systematic review work and research synthesis.} that reports directly on collected and analyzed data, e.g., a paper reporting a user study. \emph{Secondary research}, then, is one step removed: a paper that reports on a collected and analyzed sample of primary research papers: systematic reviews are one example of secondary research. We see no issue in carrying these concepts forward for research synthesis within HCI.

With these foundational concepts in hand, how then do we process the material that is primary research into the outcome that is secondary research? Unfortunately, the structure of a typical systematic review process is more contested than ideal \cite{krnicmartinic2019definition}. \citet{krnicmartinic2019definition} suggest the following components, listed in procedural order: \quoting{i) a research question; ii) sources that were searched, with a reproducible search strategy (naming of databases, naming of search platforms/engines, search date and complete search strategy); iii) inclusion and exclusion criteria; iv) selection (screening) methods; v) [a critical appraisal of] the quality/risk of bias of the included studies; vi) information about data analysis and synthesis that allows the reproducibility of the results.} \citet{haddaway2016systematic} instead compare requirements posed by institutions that promote evidence-based research through systematic reviews, e.g., the Cochrane Collaboration. They suggest three basic standards:  \quoting{%Whilst the exact format of SRs may differ between the SR-coordinating bodies (including the Cochrane Collaboration, Campbell Collaboration, Collaboration for Environmental Evidence, European Food Safety Authority, the EPPI-Centre, and the Centre for Reviews and Dissemination.), 
(i) [...] methods should be described in sufficient detail to allow full repeatability and traceability; (ii) [...] a systematic approach to identifying and screening relevant academic and grey literature, and (iii) [...] critical appraisal of the validity (quality and generalisability) of included studies to give greater weight to more reliable studies.}
%For the purpose of positioning this alt.chi paper, we wish to put forth a more comprehensive set of several noteworthy aspects that---from our perspective---should be included or considered for a definition: 
With the plurality of our field in mind, we draw out the following general characteristics:
(i) an \emph{a priori} developed and pre-registered protocol i.e., full documentation of the planned review procedure, as well as clearly and comprehensively articulated research questions, search processes, screening processes, data extraction processes, and the means of quality appraisal (or a rationale for its omission); (ii) data analysis and synthesis methods; and (iii) a discussion that transparently addresses limitations in both search and synthesis, and for both method choices and results. 

The components necessary for a review to be ``systematic'' remains an open question in HCI. Is pre-registration necessary, especially if a similar registration already exists? Does every review require an assessment of quality of the primary research? Are certain tools or platforms required, such as the use of the ACM Digital Library or IEEE Xplore, which are foundational for primary and secondary research publishing but not without their quirks and outright glitches? Further, the specifics of \emph{how} the steps should be conducted in practice are similarly unclear and in disarray. As one example in HCI, there is no consensus or even weighing in on the trade-offs for the choice between single and double screening---is one person's decision enough, or is at least one other required? Can the other(s) simply review rejected items, i.e., to avoid false negatives? Can the work be divided up between different people? Should there be a "storming and norming" process to get people on the same page or even some form of inter-rater reliability metric? Do we let go of generalizability and accept epistemological diversity? Should we adopt the aspirations to be practical and flexible and simply transparent, as advocated by \citet{braun_reflecting_2019} in their \emph{reflexive} thematic approach? On that note, \emph{can} we meaningfully and appropriately draw from other methodologies to inform research synthesis? These are just some of many methodological questions that researchers in other fields have been exploring in recent years~\cite{mahtani2019single,gartlehner2020single}, yet each one alone already raises an array of questions and provocations for the context of HCI research.

\epigraph{\textbf{Systematic Review.} \texttt{\textipa{/sIst9'matIk rI 'vju:/}}. \texttt{<<DefinitionError: term 'systematic review' is not defined>>}.}{\textit{Human-Computer Interaction, Probably}}

The systematic review in its modern form can primarily be traced back to the medical field, where the goal is to synthesize the results of multiple randomized controlled trials to better estimate the effect of a specific intervention~\cite{starr2009origins}. When the effect sizes in very similar studies are synthesized via statistical methods, it is considered a systematic review with meta-analysis \cite{cooper2015research}. The term ``meta-analysis'' is sometimes used in HCI to refer to review work without statistical aggregation of effect sizes, presumably in a more literal interpretation of the term ``meta'' to account for a paper that reports on one or more analyses, e.g., \cite{colley2022accessibiliity, volkel2020what}.
Other fields have adjusted synthesis methodology or created their own to suit their needs, for example fields and subfields that do not conduct (m)any randomized controlled trials \cite{tong2012enhancing, topor2020integrative}. This parallel methodological evolution in multiple fields has led to a dizzying array of closely related but different synthesis methods and review types: scoping review, rapid review, mapping review, review of reviews, (best-fit) framework synthesis, mixed-method synthesis, among many others \cite{sutton2019meeting}. Uptake of these methods as well as guidelines for their usage varies wildly, as do opinions on which of these are or are not ``systematic.'' As these fields have matured, they have started to face another flood of papers, this time with secondary rather than primary research \cite{ioannidis_mass_2016}. The waterfall does not stop at the pond, but cascades ever further: for a while already, academic research literature has featured tertiary research 
\cite{slim2022umbrella,aromataris2015summarizing,dasilva2011six} and even occasional examples of ``quarternary'' research \cite{mentis2021nongenetic}. 
We have no reason that this will not be the case for HCI as well; %but, from under the shadow of the titans, 
now is the time to act and seek a new path forward.
%\katja{Note to self: Need to integrate \citet{snyder2019literature} etc}

\fancyOrnament

\section*{Input Coordinates: Tracing Out Open Questions and Countering Objections}
\lettrine{T}{he} alt.chi website expects submissions to be \quoting{controversial, risk-taking, and boundary pushing} \cite{altchiwebsite}---so why are we writing about systematic reviews, when they are an established methodology, even a gold standard, that can highlight existing knowledge (``backward-looking'' \cite{niiniluoto2019scientific}) as well as create new forms of knowledge from what came before (``forward-looking'' \cite{niiniluoto2019scientific})? 
Surely this is not a controversial topic? 
Yet somehow it is: in our own experience when submitting and reviewing papers in HCI, we have come across a broad range of expectations and opinions about:
\begin{itemize}
    \item whether systematic reviews, as a form of secondary research that heavy relies on primary research, has a place in HCI, since such work does not always lead to a novel outcome in the traditional sense;
    \item what systematic reviews are for \emph{(providing an objective and comprehensive overview of a subfield vs. providing an opinionated narrative vs. providing an estimation that answers a very specific question; establishing consensus vs. providing a subjective but substantiated perspective)},
    \item how they should be conducted \emph{(based on a range of specific guidelines; ignoring or including qualitative research; with or without meta-analysis; with or without critical appraisal or double screening or data extraction forms or ...)}, 
    \item what forms of knowledge they can and should produce \emph{(``maps'' vs. synthesized effect size estimates vs. taxonomies, theories or frameworks vs. new research questions and directions vs. new primary research or instruments or prototypes)}, and
    \item basic terminology and definitions \emph{(when should a review be considered systematic; what is a meta-analysis; etc.)}
\end{itemize}
%\lettrine{O}{r:} \emph{It Shouldn't Be ``controversial, risk-taking, and boundary pushing'' \cite{altchiwebsite}, But Somehow It Apparently Is.} \\

Let us invoke an imaginary HCI researcher, who sees no benefit to systematic reviews and considers them procrustean:

%\begin{quote}
%    \textbf{Procrustean.} \texttt{\textipa{/pr9(U)'kr2stI9n/}}. Of, relating to, or resembling the practices of Procrustes (see Procrustes n.); (hence) enforcing uniformity or conformity without regard to natural variation or individuality.
%\end{quote}
\epigraph{\textbf{Procrustean.} \texttt{\textipa{/pr9(U)'kr2stI9n/}}. Of, relating to, or resembling the practices of Procrustes (see Procrustes n.); (hence) enforcing uniformity or conformity without regard to natural variation or individuality.}{\textit{Oxford English Dictionary}}

As we have outlined above, there are reasons to come into such a position within HCI, so we give this perspective a platform and trace out likely concerns. % with curiosity and empathy.
This researcher might reasonably ask: Will systematic reviews lead to no one reading the original papers anymore? 
We again emphasize that it is generally not possible to stay up to date and read all papers in HCI. Sorry. That ship has sailed. 
Yet it may be too simplistic and disillusioned to respond that "nobody reads anything anyway"---even though it seems that we do not engage with cited work as critically and comprehensively as we should \cite{marshall2017throwaway}. 
Systematic reviews \emph{could} indeed shift or divert citations from primary research papers. Reviews are easy to cite for general overview purposes, and without systematic reviews, the same authors might cite a couple of hand-picked primary research papers instead. However, researchers tackling a particular topic or carrying out work within the same domain will still cite the most relevant papers directly---or should.
Still, we acknowledge that an increase systematic reviews might affect citation practices, especially if we consider \emph{who} is writing them (and who is not) as well as \emph{how}: \quoting{citations have politics}~ \cite{citationaljusticecollective2021following}. As noted by \citeauthor{kumar2021braving}: \quoting{How work is written about also matters because it can distort or even erase contributions over time} \cite{kumar2021braving}. 
However, a well-conducted systematic review should gather and give platform to a broad and unbiased selection of papers grounded in a comprehensive search strategy and self-reflective quality assessments. It could thus help to reduce biases in how we cite and pay attention to existing research, i.e., be self-correcting in the same spirit as the scientific method. Rather than encouraging us to cite what (or who) we know, which may not represent the diversity of the field but rather our social networks  \cite{gallotti_effects_2019}, systematic review procedures can broaden our horizons and create greater inclusion in citation practice. Further, a well-conducted systematic review is itself a form of in-depth critical reflection and engagement with the primary research in its corpus. While it may ``steal'' some citations, it should itself cite the primary work and likely also elicit future citations for it going forward\footnote{We might also question when citations are truly meaningful or useful, given that they can just as much indicate social power differentials as scholarly engagement. Systematic reviews could help us dodge our natural inclination as social animals towards popularity metrics, %and also false cues to "trust" certain papers over others based on social metrics like popularity a
as operationalized in citation counts.}. 
%We again emphasize that it is generally not possible to stay up to date and read all papers in HCI. Sorry. That ship has sailed. 

Our imaginary researcher might next ask: Will systematic reviews lead us to enforce a procrustean norm in our synthesized results that entirely ignores all the beautiful variations in each of the individual papers?
This may be true. But maybe such variation does not always help us with our goal in the moment. When seeking a good (enough) answer to a specific question based on the field's currently available research, perhaps those variations are not always useful or relevant at a meta level. %Further, some (qualitative) synthesis methods take into account detail and nuance in individual papers quite extensively. 
In fact, extending the metaphor of Procrustes to user studies can show why these objections should not be an issue. As a field of study, we generally do not shirk the individual user when drawing on the results of a n=30 user study to infer how it might work for the user group as a whole. By posing implications and conclusions about a specific research question based on a single user study, we are not aiming to define or enforce a norm that ignores the beautiful variations of the individual participants. Rather, we offer a slice of the available experience with the resources at our disposal and then turn to other methods to explore and showcase the variations we could not get to in one study. Similarly, we can combine systematic reviews with specific methods of analysis to draw conclusions, and gain nuance and rich, situated understanding.

Finally, we turn towards \citeauthor{blackwell2015hci}'s perspective on HCI as a field of study:  Perhaps the goal of HCI should \emph{not} be to \quoting{develop and maintain a stable body of knowledge, but rather to be the catalyst or source of innovation}. This would instead require that we as HCI researchers engage in 
%joyously, if at times uncomfortably, find ourselves engaging in 
scholarship that is \quoting{questioning, provocative, disruptive and awkward}~\cite{blackwell2015hci}. This could be an argument against systematic reviews, as the goal of synthesis is often to stabilize and find firm ground in the shifting sands of our field. Still, \citet{blackwell2015hci} also emphasize the importance of \quoting{reflective practice}---which itself is something that knowledge synthesis through systematic reviews can deliver and structure. We suggest that systematic reviews, with all of their own methodological diversity, have the potential to be part of both the development of stable ground \emph{and} disruptive practice within knowledge production in HCI. 
%\citet{blackwell2015hci}: \quoting{What if the purpose of HCI were not to develop and maintain a stable body of knowledge, but rather to be the catalyst or source of innovation? Perhaps the function of HCI is to be questioning, provocative, disruptive and awkward in relation to other disciplines - particularly in relation to those disciplines that underlie our professional
%affiliations.} yet also \quoting{I suggest that these are the key skills of the HCI community: willingness to engage with technical practice and the desire to make interventions. Doing so in a way that is consistently effective and innovative requires both collaboration and reflective practice.} - systematic reviews can be PART of this disruptive practice

\fancyOrnament

\section*{Charting a New Trajectory: Critical Issues and Provocations}
\lettrine{B}{ased} on our experiences of planning, conducting, and publishing several systematic reviews (as well as some less than systematic ones, according to our current understanding of the term), we here present critical issues and provocations that HCI as a field needs to grapple with in the journey to answer the questions and calls raised by this manifesto. These concern the knowledge-building ecology surrounding primary and secondary research in HCI as shaped by the epistemological diversity within the HCI research community. 
However, we also foresee friction in supporting secondary research within the HCI publishing ecosystem (expectations and requirements of conference proceedings and journals, their reviewers, editors/associate chairs, and subcommittees) and technical infrastructure (e.g., the available and relevant databases and their search functions)---see \autoref{fig:ecosystem}. We map these out next.

\pgfdeclarelayer{background}
\pgfsetlayers{background,main}

\definecolor{col1}{RGB}{102,205,170} 
\definecolor{col2}{RGB}{0,128,128}

\begin{figure}[t]
{
\vspace{-.95cm}
\footnotesize
\resizebox{.4\textwidth}{!}{
\begin{tikzpicture}
    \tikzstyle{block}=[
        draw,
        node distance=0.5cm,
        rounded corners,
        thick,
        text width=.7cm,
        minimum height=1.6cm,
        minimum width=1cm,
        align=center,
        font=\tiny,
        every node/.style={inner sep=0,outer sep=0},
        fill=col1,
    ]
    \tikzstyle{block2}=[
        draw,
        dashed,
        node distance=0.5cm,
        rounded corners,
        thick,
        text width=1cm,
        minimum height=2.5cm,
        minimum width=4.33cm,
        align=center,
        anchor=east,
        font=\tiny,
        every node/.style={inner sep=0,outer sep=0},
    ]

    \tikzstyle{header}=[rotate=90, align=center]
    \tikzstyle{line}=[draw, thick]%-latex for arrow lines

    \node[block2,fill=col1, opacity=0.2] (infra) {};
    \node[anchor=center, text width=2.8cm, rotate=90] at ([xshift=1.85cm, yshift=.5cm]infra.center) {\tiny technical infrastructure};
    \node[block2, fill=col2, opacity=0.2, minimum width=3.7cm, minimum height=3cm, anchor=north] at ([xshift=1.86cm]infra.north west) (eco) {};
    \node[anchor=north] at ([yshift=.45cm]eco.south) {\tiny publishing ecosystem};

    \node[block, minimum width=1.5cm] at ([xshift=1.15cm]infra.west) (pr) {};
    \node[anchor=north] at ([yshift=.45cm]pr.center) {\tiny primary research};
    \node[block, right = of pr, minimum width=.7cm, dotted] (sr) {secondary research};
    \node[circle, draw, dashed, below = of pr.center, yshift=.9cm, minimum size=10pt] (circle1) {};
    \node[circle, draw, dashed, left = of circle1, xshift=1.1cm, yshift=-.1cm, minimum size=10pt] (circle2) {};
    \node[circle, draw, dashed, right = of circle1, yshift=-.1cm, xshift=-1.1cm, minimum size=10pt] (circle3) {};

    % Draw edges
    \path[line,->] (pr) -- (sr);
    \path[line,->] (pr.30) -- ([xshift=.5cm]pr.30);
    \path[line,->] (pr.330) -- ([xshift=.5cm]pr.330);

\end{tikzpicture}
}}

\caption{The academic ecosystem 
%involved in the systematic review-lution 
consists of primary research based on a broad range of different methods and research paradigms, which can then be synthesized in secondary research. All research is created 
%and represented within, as well as often 
and shaped by the academic publishing ecosystem---e.g., venues, reviewers, and conference (sub)committees---and technical infrastructure---e.g., databases and their search functions.}\label{fig:ecosystem}
\Description{A schematic of the academic ecosystem involved in the systematic review-lution, consisting of two main components: a primary research rectangle, and a secondary research rectangle. Arrows point from the primary research component to the secondary research in which it is synthesized. Both components are situated within a rectangle labelled as the publishing ecosystem, and at the same time, within another labelled as the technical infrastructure.}
\end{figure}
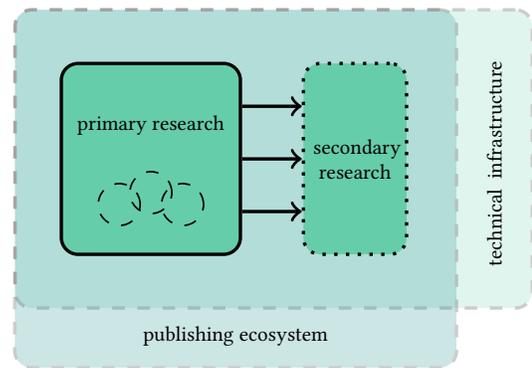

\paragraph{Primary Research Reporting}
The foundation of research synthesis generally, and systematic reviews specifically, is built upon the reporting of primary research. Yet empirical work in HCI---viewed across the field as a whole---is disparate; how we report is varied and sometimes spotty. 
This is certainly not a novel criticism; we echo calls from other researchers who are critically reflecting on our research reporting practices, e.g., with regards to the reporting of race and ethnicity data \cite{chen2022collecting}, brain signal experiment data \cite{putze2022understanding}, participant compensation data \cite{pater2021standardizing}, inter-rater reliability in qualitative research \cite{mcdonald2019reliability}, specific measures \cite{seaborn_measuring_2021} and questionnaires \cite{law2018systematic, hughes2023growing}, engagement with self-determination theory \cite{tyack2020self}, artifact descriptions \cite{gerling2022reflections}, and inferential statistics \cite{cairns2007hci}, to list just a few. % \cite{putze2022understanding,law2018systematic, hughes2023growing,pater2021standardizing,chen2022collecting,gerling2022reflections,mcdonald2019reliability}, to list just a few)
These issues may arise in part due to page limits or efforts to ensure paper length matches perceived contribution, but may also be due to lack of community-driven standardization and education. 

This complicates research synthesis in secondary research because it makes results difficult to compare and weigh. Again, we recognize that this may not always be the goal, but it often is in the HCI world. Yet how can we point to what works and what does not if we cannot synthesize results with a high degree of rigour or systematicity? 
The good news is that using existing guidelines for reporting more will likely also help with secondary research simply by making the reported primary research more comprehensive and comparable. Still, it may be worth examining to what extent existing guidelines for reporting primary research can support follow-up secondary research. 

Further, given that criticism of reporting in HCI has been expounded for many years, perhaps it is time to consider pointing authors and reviewers in HCI to such guidelines more explicitly. 
There are already hints of conferences adding a bit more structure to the submission process. For example, since 2021, at least one ACM conference has required authors to indicate \quoting{the primary and secondary contribution type of their paper} (empirical-qualitative; empirical-quantitative; empirical-mixed-methods; artifacts-technical; artifacts-design; theoretical; or meta-research\footnote{Adapted from \citet{wobbrock2016research}'s classification of research contributions in HCI.}), to assist with reviewer fit to assigned papers \cite{chiplay2021website}. We could add a requirement for papers to include a structured abstract of sorts as supplementary material---tailored to the contribution type. 
This could support not only future secondary research, but also the reviewing process itself, by providing a concise overview of the conducted work that reviewers can quickly and easily digest.

\paragraph{Epistemological Diversity of CHI}
We next address the role that our different ways of knowing in HCI play in the synthesis of primary work. This needs to be considered both from the perspective of the diversity of research \emph{within} primary work in HCI, but also in the diversity of methods that we draw on for synthesizing it in secondary research. 

How do we approach any kind of formalization of systematic reviews for as broad a field as HCI? 
Research in other fields has in recent years looked at the methodology of mixed methods reviews in more detail \cite{hong2017convergent, sandelowski2006defining}. 
%Based on research design models for mixed methods studies, \citet{sandelowski2006defining} distinguish mixed methods synthesis methods as \quoting{segregated} (separately analyzing the different types of evidence), \quoting{integrated} (using a combined mixed-methods analysis and synthesis that typically requires data transformation), or \quoting{contingent} (sequential syntheses, with each one informing the next one). 
%In reviewing reviews of mixed methods studies, \citet{hong2017convergent} found that these can approach the review in a convergent or sequential manner: the quantitative and qualitative data is either analyzed in a parallel fashion (\quoting{convergent}), or one after the other, with the first analysis informing the second one (\quoting{sequential}). When the approach is convergent, the reviews either use the same synthesis method for both types of data (\quoting{data-based convergent}); use separate synthesis methods and then integrating the results with another analysis and synthesis method (\quoting{results-based convergent}); or use separate synthesis methods and then integrate results in narrative form in the discussion (\quoting{parallel-results convergent}). 
To conduct a review that accommodates research results from quantitative as well as qualitative methods, we can point to the Joanna Briggs Institute (JBI) guidelines for mixed methods systematic reviews \cite{lizarondo2020jbimmsysreviews} as a starting point that covers some approaches. %mixed methods research models. %the integrated and segregated approaches. 
Currently there are only a few reviews in HCI that use these or related guidelines. Still, we think it deserves more attention from our mixed methods-inclined field of study and could greatly benefit the way we do synthesis. 

But HCI also features approaches that are situated more in design research methods, like participatory design and research through design. \citet{wolf2006dispelling} describe the field as featuring as an \quoting{inherent tension [that is] reflected in the distinctive practices and disciplinary orientations of engineering and creative design.} We agree and also highlight that this \quoting{is not an insurmountable conflict [...] both perspectives are valid} \cite{wolf2006dispelling}. 
Excluding the knowledge created through design research methods when we do synthesis decentres a significant section of our field and prevents us from accessing a truly full picture of HCI practice.  
However, to our knowledge, there are currently no methods or guidelines designed to handle and synthesize evidence and knowledge created through design research methods. We may have to develop new methods to integrate this work into systematic reviews.  
We call on researchers familar with each approach: \quoting{any notion of rigour has to be developed within a 'firm understanding of the particular purpose of each approach'} \cite{frauenberger2015in} (\citet{frauenberger2015in} citing \citet{fallman2010establishing}).

There is no short supply of work in these fields for exploring what rigour means within different HCI approaches and how to evaluate such work for synthesis purposes. For example, \citet{wolf2006dispelling} outlines qualities in design praxis that aim to achieve \quoting{design rigor}, among them the design critique: \quoting{a designer's reflective, evaluative and communicative explanation of her design judgments and the activities in which she has engaged}. Similarly, \citet{zimmerman2007research}'s criteria or lenses for evaluating interaction design research (process, invention, relevance and extensibility) may be a useful tool for synthesis. 
For approaches like participatory design, \citet{frauenberger2015in} write about how traditionally positivist understandings of rigour need to be re-interpreted: \quoting{accountability and rigour in a post-modern scientific context is delivered through debate, critique and reflection}, and make the case for \quoting{acknowledging different ways of knowing} \cite{frauenberger2015in}. We extend this argument: not only do we need to acknowledge these different ways of knowing, we need to develop methods of synthesizing and integrating different ways of knowing, as well. 

%\katja{Integrate: \quoting{Too often, literature reviews are simply descriptive summaries of research conducted between certain years, describing such information as the number of articles published, topics covered, citations analyzed, authors represented, and perhaps methods used, without conducting any deeper analysis. While there are likely times when this can be valuable, this is not usually the case and they are not likely to be published in any journal. In truth, review articles that are a medley of word clouds and citation analyses are highly unlikely to be published. This is a pity, as these researchers have gone through the tedious work of collecting many articles without actually analyzing them in any meaningful way, and because of this, they fail to make a significant contribution.} \cite{snyder2019literature}.}
%

\paragraph{Secondary Research Reporting}
Perhaps the closest match for existing methodological guidelines towards which synthesis guidance in HCI could be oriented are efforts within software engineering (e.g., \citet{kitchenham2007guidelines}'s work), qualitative health research (e.g.,~\citet{tong2012enhancing}'s ENTREQ, or \citet{cooke2012beyond}'s SPIDER framework), and quite recently, \citet{topor2020integrative}'s NIRO-SR for non-intervention studies (still a preprint). However, we strongly believe that HCI will need to also draw on synthesis methods that more explicitly \emph{combine} quantitative and qualitative work: 
To quote \citet{reeves2015human}, we need \quoting{more reviews of and reflections upon the landscape of different forms of reasoning in HCI and through this better ways of managing how potentially competing disciplinary perspectives meet together.}
Guidance for pulling together evidence from different disciplines and methodologies \emph{does} exist (e.g., \cite{lizarondo2020jbimmsysreviews}) although it is rare. Yet how well this works in HCI is an open question; currently, there is essentially no uptake in our field. 

%Based on an analysis of systematic reviews at CHI (forthcoming), 
When existing systematic reviews at CHI cite a guideline for their method, they primarily reference PRISMA \cite{page2021prisma} (e.g., \cite{bergstrom2021how,hirzle2021critical,macarthur2021youre}). %\footnote{Alternatively, they reference previously published systematic reviews as the guiding example(s) in lieu of actual guidelines (e.g., \cite{velt2017survey}).}.
The PRISMA figure, specifically, is popular, as it can be a great way to illustrate the search and screening process. 
However, the PRISMA guidelines as a whole were made for reporting systematic reviews and meta-analyses of intervention studies in the medical field~\cite{page2021prisma}. Most HCI reviews---even if they state that they followed the PRISMA guidelines!---do not actually answer all (or even most) of the PRISMA checklist items \cite{page2021prisma,page2021prismaEE}. %, even though it might strengthen the paper's contribution substantially. 
For example, how often have you seen a systematic review at CHI report a quality assessment and/or risk of bias in each study\footnote{\quoting{Item 18. Present assessments of risk of bias for each included study} ~\cite{page2021prismaEE}} or certainty in the evidence as a whole\footnote{\quoting{Item 15. Describe any methods used to assess certainty (or confidence) in the body of evidence for an outcome} \cite{page2021prismaEE}, e.g., via the GRADE framework for evaluating quality of evidence in a review \cite{guyatt2008grade}.}? 
Further, because of the medical world's focus on meta-analyses, several PRISMA items are designed for statistical synthesis methods that reviews in HCI only very rarely employ (e.g., explorations of causes of statistical heterogeneity\footnote{\quoting{Item 13e. Describe any methods used to explore possible causes of heterogeneity among study results (such as subgroup analysis, meta-regression)} \cite{page2021prismaEE}}), and are thus simply not applicable to the kinds of reviews that we (can) conduct.
%because they do not apply to the kinds of review that we conduct. (Although in some cases, perhaps they should!) For example, how often have you seen a systematic review at CHI report an exploration of causes of statistical heterogeneity\footnote{\quoting{Item 13e. Describe any methods used to explore possible causes of heterogeneity among study results (such as subgroup analysis, meta-regression)} \cite{page2021prismaEE}}, risk of bias\footnote{\quoting{Item 18. Present assessments of risk of bias for each included study} ~, or certainty in the evidence\footnote{\quoting{Item 15. Describe any methods used to assess certainty (or confidence) in the body of evidence for an outcome} \cite{page2021prismaEE} e.g., via the GRADE framework for evaluating quality of evidence in a review \cite{guyatt2008grade}.} in the synthesis \cite{page2021prisma,page2021prismaEE}? 
The PRISMA figure may be useful, but the guidelines are, for the most part, not actually appropriate for our field---at least not past the search procedure when it comes to the synthesis methods at the heart of the review. 

A quick search for "systematic review" in the ACM DL shows a sharp increase in systematic reviews being produced. %(perhaps simply by way of a general increase in research output, or more recently fueled by researchers looking for alternative ways to publish when user studies were made more difficult or even impossible during the COVID-19 pandemic). 
This means that now is an important moment to \textbf{STOP} and reflect on the methods we use for systematic reviews in our field. We need to figure out what we mean when we use the term \quoting{systematic} in the context of review work, and what we expect in terms of best practices. We need to 
%stop basing our reviews on already published reviews, without 
report methods clearly and comprehensively, including how we adapted guidelines
% without actually stating which parts we conducted in the same way and which parts we adapted 
 to our own use. 
We need to look more deeply into synthesis methods and carefully choose, name, and rationalize our choices. We may also want to look into structured abstracts as supplementary materials for secondary research (for example, \citet{haddaway2018roses}'s ROSES 
%---\quoting{a pro forma and flow diagram designed specifically for systematic reviews and systematic maps in the field of conservation and environmental management}---
%offers an option for this that 
could either be borrowed directly or adapted for HCI research). 

\citet{oulasvirta2016hci} put forth that HCI needs more \quoting{conceptual contributions that link empirical findings and the design of technology} to make our research findings actionable and create \quoting{integrative types of knowledge.} We argue that by putting effort into developing and upholding guidelines and standards for review synthesis and its reporting that works for HCI specifically, we will be able to improve the conceptual contributions that HCI can make as a field. If we view HCI as a field defined by its problem-solving capacity \cite{oulasvirta2016hci}, then systematic reviews---when done rigorously---can directly help to improve several of the criteria they propose as important for problem-solving: it can help us develop a better understanding of how well solutions \emph{transfer} and inform our degree of \emph{confidence} in them.      

%Paper requirements: structured abstract as supplementary but for secondary research e.g. ROSES \cite{haddaway2018roses}? require contribution type and methodology keywords like CHI PLAY does. Do signpost it as a systematic review if it is one, and clarify why you think it is one. Don't call it meta-analysis unless you're doing statistical meta-analysis.  

%\citet{topor2020integrative}

%\citet{reeves2015human}: 
%\quoting{We will need more reviews of and reflections upon the landscape of different forms of reasoning in HCI and through this better ways of managing how potentially competing disciplinary perspectives meet together.}

%\citet{blackwell2015hci}: \quoting{}

\paragraph{Venues and Subcommittees}
\citet{marshall2017alittle} lamented that HCI has few explicit publication formats that invite critical discussion: \quoting{none of the major [venues] 
%journals such as ACM TOCHI, International Journal of Human Computer Studies, or indeed CHI conferences 
have any format for critical response to published articles [...]
%, beyond questions at a CHI presentation. There are no culture of pre-prints and early criticism. This means that 
once a piece of HCI work is in publication, it is unlikely to attract any critical discussion.} Critical discussion instead is more likely to take place in 
%Many have moved from formalized venues to other places: 
social media, Slack workplaces and Discord channels, and other unofficial venues. %, perhaps even impromptu congregations around the water bowl in the lab.
Systematic reviews could perform the function of critical discussion in a rigorous and formalized way, accessible to the community of practice as a whole. Yet there is no clear place for them, either.  
Perhaps the only publication venue in HCI that explicitly welcomes reviews (\quoting{survey papers}) is the ACM Computing Surveys (CSUR) journal, but they make no mention of systematicness in their author guidelines \cite{csur}. 
CHI as the \quoting{flagship conference of the discipline} \cite{liu2014chi} features only one subcommittee---Health---that mentions (systematic) reviews as a method in their description \cite{chi2023subcommittee}. 
Even subcommittees that describe themselves as \quoting{epistemologically pluralistic [and] welcoming of a range of perspectives, approaches, and contributions} \cite{chi2023subcommittee} can recruit associate chairs and reviewers that do not consider systematic reviews as a methodology per se and may be inclined to reject them for that reason alone. 
Reviewers in HCI as a whole have wildly different expectations and methodological expertise when it comes to reviews; a little more agreement would go a long way. 
%this is an unfortunate state of affairs that needs to be improved. 

%\citet{marshall2017alittle}: \quoting{none of the major journals such as ACM TOCHI, International Journal of Human Computer Studies, or indeed CHI conferences have any format for critical response to published articles, beyond questions at a CHI presentation. There are no culture of pre-prints and early criticism. This means that once a piece of HCI work is in publication, it is unlikely to attract any critical discussion} 

%We suspect that this might be part of the reason why some reviews end up in extended abstracts at CHI instead (e.g., as late-breaking work \cite{chen2022collecting})---even though this makes comprehensive reporting even more difficult---or in other publication venues entirely. 
A perception of systematic reviews not producing \quoting{novel} work may be a partial reason for this issue. 
%For example, a(n informal) review paper by \citet{vanturnhout2019practical} was not accepted in part due to "a lack of novelty," and was instead published as a white paper. 
For example, the TOCHI journal warns that they \quoting{rarely publish[...] survey papers unless they offer a major original contribution.} We note that reviews \emph{absolutely} can produce original contributions based on the synthesis, e.g., intermediate-level knowledge like taxonomies \cite{brudy2019cross}. When it should be considered \quoting{major}, and whether or when a systematic overview of existing work should be considered an \quoting{original contribution} is something that might be helpful for TOCHI to describe in more detail for potential authors, and indeed something that we should discuss as a field. %ndeed, the fact that we have not yet come together as a community to discuss these matters directly may have seeded and be maintaining the situation.

%meta work in general has no own subcommittee at CHI

%\citet{marshall2017alittle}: \quoting{none of the major journals such as ACM TOCHI, International Journal of Human Computer Studies, or indeed CHI conferences have any format for critical response to published articles, beyond questions at a CHI presentation. There are no culture of pre-prints and early criticism. This means that once a piece of HCI work is in publication, it is unlikely to attract any critical discussion} 

%Example of a paper that looks super relevant to the theoretical advancement of HCI (an informal review, possibly) but was not accepted in part due to novelty, published as white paper instead: \citet{vanturnhout2019practical}

\paragraph{Infrastructure: Digital Libraries and Machine Learning Approaches}
Our digital libraries are poorly documented and barely evaluated. 
Results can vary wildly over time. This is sometimes expected (i.e., numbers go up as more research is published); however, it sometimes also \emph{decreases} due to adjustments in metadata\footnote{And on some days, databases are simply buggy: on one memorable occasion, we noted the ACM DL reporting 0, then 200+, then 0, then 500+ results for the same search within a single day.}. Metadata in publication databases often has errors and cannot necessarily be relied on \cite{franceschini2016empirical, franceschini2016museum}. Additionally, what databases cover is not always entirely clear and can vary based on institutional access---e.g., the \quoting{Web of Science Core Collection} consists of different sub-data sets depending on university subscription \cite{liu2019data}. 
This makes one of the fundamental goals of systematic reviews---namely, that it should be possible to reproduce the results---rather difficult. It is considered best practice in other fields to conduct searches on multiple databases. Perhaps we need to consider doing multiple searches over several days to try to mitigate database fluctuation. However, perhaps we also need to re-consider or be clearer about what we require for a systematic review to be "reproducible": What do we mean when talk about reproducing results?
For example, as long as the search queries themselves are reported, and the records of papers found in each step%(initial, interim, final) are listed%(initial search results, interim results after each screening stage, final corpus)
, then perhaps we should not require the search to yield the same number of results, simply because we cannot rely on the databases to be consistent.

Still, there are additional issues with designing multiple searches to be \emph{comparable} across databases. Databases use a variety of different keyword and filter options, and often they are only poorly documented. Guidance for creating comparable searches across, for example, ACM and Scopus, would be highly beneficial for synthesis in our field. Current ACM DL tutorials are not sufficient for this purpose, and contacting the ACM DL team about database and search specifics has been unproductive. %design comparable searches across databases. 
%We have contacted the ACM DL team multiple times about specific errors in results, and 
%more general questions about 
%how the database's search functions work in detail---without success.
%For the more specific questions, the ACM DL contact person told us they needed to contact the team who had implemented the database, yet despite multiple emails to follow up, we never received an answer. For the more general questions, they referred to the 
%ACM DL tutorials---yet these are not nearly sufficient to be able to design comparable searches across databases. 
One option is to work in concert with publishers in % and other stakeholders who may see the value in review work and simply do not know how to support it in the design of these systems. We can imagine 
a participatory design project with ourselves as the target "end-users" directing the design of these systems in a more fruitful direction for supporting review work. %There may even be opportunities for "in the wild" work, where we can (re)run the types of search procedures we wish to undertake and compare database output---queries and the display of items on the UI---while the system designers iterate in response to our "just-in-time" and contextualized feedback. 
Another option to consider is automation. With the growth in artificial intelligence and machine learning, the landscape of digital infrastructure surrounding databases and publication searches now also features tools for (semi-)automated search (e.g., Research Rabbit\footnote{\url{https://www.researchrabbit.ai/}, last accessed: 23 Nov, 2022}) or 
 screening (e.g., ASReview\footnote{\url{https://asreview.nl/}, last accessed 15 Dec, 2022}). These may be of interest for reviewing the field, but to what extent they can and should be used in formal systematic reviews is an open question---especially as the exact data sources and how often they are updated is often not made explicit. Perhaps a participatory design approach can again be useful as a starting point. %where the foundational system is realized based on essential user needs and use cases that can then be customized through automated means during actual use.
%What about tools like Research Rabbit etc? Unclear what data source and how often updated---how well can we use this now for actual synthesis and formal systematic literature reviews? 

Finally, we note that there is little information on what kinds of publications relevant to HCI are found within which databases. \citet{gusenbauer2022search} created a discipline-based coverage map of a wide range of academic databases, giving us a first hint. However, HCI was not included in this disciplinary coverage map; it may be worth creating a disciplinary coverage map of databases for HCI specifically. This would give us a better idea of what kind of HCI research can be found in which database, and provide guidance on which databases to chose for specific research questions.

\fancyOrnament

\section*{Engage: A Call to Action for Research Synthesis in HCI}

\lettrine{A}{s} a research community, we need to come together %, regardless of expertise or experience with systematic review work or other forms of research synthesis, 
and decide what actions need to be taken towards building a set of standards that is rigorous yet inclusive of the diversity of work that we do in HCI. We do not aim to be prescriptive in this manifesto, but we do offer some ideas for what to aim for based on the discussion so far:

\begin{itemize}
    \item a shared understanding of what should be considered a systematic review, the desired and possible outcomes of systematic reviews, and the forms that systematic reviews can take when exploring diverse evidence resulting from different research paradigms (quantitative, qualitative, mixed-methods, as well as design research methods) 
    \item a shared understanding of what best practices we want to encourage in secondary research methods: double screening, extraction, critical appraisal, protocol development and preregistration, etc., specifically through an agreement on standards (e.g., for critical appraisals of primary research: what kind and when)
    \item unearthing how the digital libraries relevant to HCI work (e.g., query filters) and what they cover
    \item better infrastructure in our publishing ecosystem: is it time for a subcommittee or track for research synthesis and meta science? Should we require structured abstracts or checklists for primary and/or secondary research? %For secondary research in particular, such an abstract might be a lot more useful than a supplementary video.    %do we need a replication SC? (HRI already started: open science track). we don't know where this should go but we need this SPACE. 
    \item robust descriptions of and/or access to the interactive artifacts reported on in primary research papers to support research synthesis about them
    \item exploration of the design and use of living reviews \cite{elliott2017living}---as interactive systems, HCI expertise could be particularly beneficial here
\end{itemize}

%\fancyOrnament

%\paragraph{Proposed Presentation Format}
Our goal is to begin a discussion and gather different experiences and opinions of researchers on the role that systematic reviews should play, on what a systematic review should look like, and how systematic reviews are currently valued and received within the CHI community---and more broadly, within HCI as a whole.

\fancyOrnament

\begin{acks}
Many thanks to Maximilian Altmeyer for feedback on an earlier draft of this alt.chi paper, and to all of our colleagues for the many lively discussions on these topics over the years.
\end{acks}

\bibliographystyle{ACM-Reference-Format}
\bibliography{bibtex}

\end{document}